\documentclass[pre,aps,showpacs]{revtex4}
\usepackage{graphicx}

\begin{document}
\title{Chaos in temperature in the Sherrington-Kirkpatrick model}

\author{T. Rizzo}
\email{tommaso.rizzo@inwind.it}
\affiliation{Dipartimento di Scienze Fisiche, Universit\`a ``Federico II'',
         Complesso Monte S. Angelo, I-80126 Napoli, Italy}

\author{A. Crisanti}
\email{andrea.crisanti@phys.uniroma1.it}
\affiliation{Dipartimento di Fisica, Universit\`a di Roma ``La Sapienza''}
\affiliation{Istituto Nazionale Fisica della Materia, Unit\`a di Roma, 
             P.le Aldo Moro 2, I-00185 Roma, Italy}

\begin{abstract}
We prove the existence of chaos in temperature in the Sherringhton-Kirkpatrick model. 
The effect is exceedingly small, namely of the ninth order in perturbation theory.
The equations describing two systems at different temperatures constrained to have a fixed overlap are studied analytically and numerically, yielding information about the behaviour of the overlap distribution function $P_{T_1,T_2}(q)$ in finite-size systems.
\end{abstract} 


\maketitle

\section{introduction}

In this work we address the question of chaos in temperature in the Sherringhton-Kirkpatrick (SK) model. According to the Parisi solution \cite{MPV} the SK model has many equilibrium states at a given temperature. These states are correlated: the overlap between different states can take a whole range of values from zero to the self-overlap $q_{EA}$. A natural  and simple question is:
what is the overlap between states at different temperatures? 
According to the chaos in temperature hypothesis the overlap between any couples of equilibrium states at different temperatures is zero. 
Consequently, the probability distribution of the overlap of configurations equilibrated at different temperatures $P(q)$ should be a delta function centered at $q=0$. One can be more quantitative 
considering the convergence to a delta function of the function $P_N(q)$ corresponding to a finite system size $N$. Standard arguments tell us that for large $N$ this quantity is given by the large-deviations formula
\begin{equation}
P_N(q)\propto \exp[-N \Delta F(q)]
\label{PNQ}
\end{equation}
the quantity $\Delta F(q)=F(q,T_1,T_2)-(F(T_1)+F(T_2))$ is the free-energy cost, {\it i.e.} the  difference between the free energy of a system of two copies at different temperatures of a spin-glass with the same realization of the disorder constrained to have a given value $q$ of the overlap, and the sum of the free energies at temperatures $T_1$ and $T_2$ (actually here and in the following we call {}``free energy{}" the opposite of the logarithm of the partition function, without the usual $\beta$ factors).
Consequently we can talk of a small or large chaos effect depending on the function $\Delta F(q,T_1,T_2)$; 
if $\Delta F(q,T_1,T_2)$ is small the convergence with $N$ of the $P_N(q)$ will be slower and the chaos effect will be seen only when considering large-size systems. In particular if $\Delta F(q)$ turns out to be zero for a finite range of values of the constraint there is no extensive chaos; furthermore a null $\Delta F$ is expected if the $P(q)$ is definitely different from a delta centered on zero. In the following we shall compute the quantity $\Delta F(q)$.

In a recent paper \cite{rizzo1}, one of us reexamined the problem through a perturbative expansion in the replica framework, showing that chaos, if at all present, should be a very small effect. 
Actually, this is a quite old problem \cite{BY,PAR,KON1,KON2,FN,RIT}, and has gained new attention in connection with the experimental effects of memory and rejuvenation (see \cite{BDHV} and references therein).  
Recently some interesting results have been found for other models. 
Sales and Yoshino \cite{SALYOSH} have convincingly shown that there is chaos in the directed polymer in random media (DPRM) in 1+1 dimensions due to the interplay between energy and entropy, a mechanism which had been postulated in the droplet theory for finite dimensional spin-glass \cite{FH}. On the other hand the existence of high correlations between systems at different temperatures has been proved in a well known class of mean-field spin-glass models \cite{rizzo2}. Some evidence against chaos in the naive TAP equations has been found in \cite{PPM}, a result which has been questioned in \cite{MS}.
Powerful numerical efforts have been devoted by Billoire and Marinari (BM) to investigate chaos in temperature  in the SK model  \cite{MARBILL1,MARBILL2}; even though they  don't actually see the effect, there  is some evidence that it may be extremely weak, thus observable only on very large lattices.

\section{The variational equations}
In the replica framework one obtains a variational expression for the free-energy functional	$F(q_c,T_1,T_2)$ of two systems with a constraint $q_c$  \cite{FPV1,FN,rizzo1}. We have expressed $F(q_c,T_1,T_2)$ in terms of a Parisi differential equation. The variational parameters are three $n\times n$ matrices $Q_1 \equiv (0,q_1(x))$, $Q_2 \equiv (0,q_2(x))$, and $P \equiv (p_d,p(x))$:
\begin{eqnarray}
F(q_c,T_1,T_2) & = & -{\beta_1^2 \over 4}-{\beta_2^2 \over 4}+{\beta_1^2 \over 4}\left(-\int_0^1q_1^2(x)dx+2 q_1(1)\right)+{\beta_2^2 \over 4}\left(-\int_0^1q_2^2(x)dx+2 q_2(1)\right)+
\nonumber
\\
& + &{\beta_1\beta_2 \over 2}\left(p_d^2-\int_0^1p^2(x)dx\right)-{\beta_1\beta_2 \over 2}(p_d-q_c)^2-f(0,0,0)
\label{F}
\end{eqnarray}
The function $f(x,y_1,y_2)$ is defined through the following differential equation:
\begin{eqnarray}
-\frac{\partial f}{\partial x}& = & {\dot{q}_1 \over 2}\left({\partial^2 f \over \partial y_1^2} +x\left({\partial f \over \partial y_1} \right)^2 \right)+{\dot{q}_2 \over 2}\left({\partial^2 f \over \partial y_2^2} +x\left({\partial f \over \partial y_2} \right)^2 \right)+
\nonumber
\\
& + & \dot{p}\left({\partial^2 f \over \partial y_1 \partial y_2} +x{\partial f \over \partial y_1} {\partial f \over \partial y_2}  \right)
\label{eqF}
\end{eqnarray}
With initial condition:
\begin{equation}
f(1,y_1,y_2)=\ln [2 e^{\delta}\cosh [\beta_1 y_1+\beta_2 y_2]+2 e^{-\delta}\cosh[\beta_1y_1-\beta_2 y_2]]
\label{iniF}
\end{equation}
\begin{equation}
\delta=\beta_1\beta_2(p_d-p(1))
\end{equation}
Notice that the initial condition depends on the difference $p_d-p(1)$.
In order to maximize the free energy functional Lagrange multipliers $P(x,y_1,y_2)$ are used much as in the standard single system problem \cite{SD,rizcris}, then introducing the following quantities
\begin{eqnarray}
m_1(x,y_1,y_2)={1 \over \beta_1}{\partial f \over \partial y_1} & ,
&
m_2(x,y_1,y_2)={1 \over \beta_2}{\partial f \over \partial y_2}
\label{defm}
\end{eqnarray}
and the operator
\begin{eqnarray}
\Omega & = & \frac{\partial }{\partial x}+ {\dot{q}_1 \over 2}\left({\partial^2  \over \partial y_1^2} +{2x\over T_1}m_1{\partial  \over \partial y_1}\right)+{\dot{q}_2 \over 2}\left({\partial^2  \over \partial y_2^2} +{2x\over T_2}m_2{\partial  \over \partial y_2}\right)+
\nonumber
\\ 
& + & \dot{p}\left({\partial^2 \over \partial y_1 \partial y_2} +{x\over T_1}m_1{\partial  \over \partial y_2}+ {x\over T_2}m_2{\partial  \over \partial y_1} \right) \ ,	
\end{eqnarray}
the saddle-point (SP) equations for the constrained systems read:
\begin{eqnarray}
\frac{\partial P}{\partial x} & = & {\dot{q}_1 \over 2}\left({\partial^2 P \over \partial y_1^2} -{2x\over T_1}{\partial  \over \partial y_1}(P m_1)\right)+{\dot{q}_2 \over 2}\left({\partial^2 P \over \partial y_2^2} -{2x\over T_2}{\partial  \over \partial y_2}(P m_2)\right)+
\nonumber
\\ 
& + & \dot{p}\left({\partial^2 P\over \partial y_1 \partial y_2} -{x\over T_1}{\partial  \over \partial y_2}(P m_1)-{x\over T_2}m_2{\partial  \over \partial y_1}(P m_2) \right)
\label{SP1}
\end{eqnarray}
\begin{eqnarray}
\Omega m_1=0 & ; & \Omega m_2=0
\label{SP2}
\end{eqnarray}
\begin{equation}
q_1(x)=\int dy_1dy_2 P m_1^2;
\ \ \ 
q_2(x)=\int dy_1dy_2 P m_2^2;
\ \ \ 
p(x)=\int dy_1dy_2 P m_1m_2;
\label{SP3}
\end{equation}
The initial conditions for the functions $m_1(x,y_1,y_2)$, $m_2(x,y_1,y_2)$ can be obtained from (\ref{iniF}) and (\ref{defm}); for $P(x,y_1,y_2)$ we find
\begin{equation}
P(0,y_1,y_2)=\delta(y_1)\delta(y_2)\ .
\end{equation}
The presence of a constraint $q_c$ leads to the following equation 
\begin{equation}
q_c=\int dy_1dy_2P(1 ,y_1,y_2)\frac{ e^{\delta}\cosh [\beta_1 y_1+\beta_2 y_2]-e^{-\delta}\cosh[\beta_1y_1-\beta_2 y_2]}{ e^{\delta}\cosh [\beta_1 y_1+\beta_2 y_2]+ e^{-\delta}\cosh[\beta_1y_1-\beta_2 y_2]}\ .\label{eqq_c}
\end{equation}
If $\delta=0$ the initial condition (\ref{iniF}) separates and eq. (\ref{eqq_c}) implies $p(1)=q_c$. 
Finally, we recall that the derivative of the free energy with respect to the constraint can be expressed as 
\begin{equation}
\epsilon={\partial F \over \partial q_c}=\beta_1\beta_2(p_d-q_c) \ .
\label{PEPSI}
\end{equation}

\section{The $\beta_1=\beta_2$ case}

When $T_1=T_2$ there are  two possible situations depending on the value of the constraint $q_c$.
For all values of $q_c$ inside the support of the overlap distribution function $P(q)$ we have solutions with $\delta=0$ and therefore $\epsilon=0$. 
As a consequence for $0\leq q_c \leq q_{EA}$ we have $\Delta F=0$ as it is to be expected since the function $P(q)$ is non trivial.
 The solutions of the SP equations satisfy $p_d=q_c$ and are:
\begin{eqnarray}
 q_1(x)=q_2(x)=p(x)=q_{Parisi}(2x) & 0\leq x\leq \frac{1}{2}x_{Parisi}(p_{d})\nonumber
 \\
 q_1(x)=q_2(x)=p(x)=p_{d} &
 \frac{1}{2}x_{Parisi}(p_{d})\leq x\leq x_{Parisi}(p_{d})\nonumber
 \\
\label{eqts}
 q_1(x)=q_2(x)=q_{Parisi}(x);p(x)=p_{d}& x_{Parisi}(p_{d})\leq x\leq 1
\label{soliso}
\end{eqnarray}
These solutions were first proposed in \cite{FPV1} for the truncated model. In \cite{rizzo2} it has been proven that they exist in any models with Replica-symmetry breaking (RSB) by noticing that they are a permutation of the standard Parisi solution. 

When $q_c$ lies outside the support of the function $P(q)$ there are  off-equilibrium solutions of the SP equations with $\delta\neq 0$ and $q_1(x)=q_2(x)=p(x)$ \cite{FPV1}. These solutions can be obtained by solving the SP  equations (\ref{SP1},\ref{SP2},\ref{SP3}) perturbatively in $\delta q=q_c-q_{EA}(T)$ and $\tau= T-T_c$. This has been done at the lower orders in \cite{FPV1} where it was found that near the critical temperature the free-energy difference grows with $\delta q^3$ in the off-equilibrium region. We pushed the computation to very high orders and we can safely claim that there are no $\delta q^2$ terms in the expansion. Recently this result was  confirmed numerically \cite{BFM}.
The off-equilibrium phase can be described also by considering what happens for $q_c\simeq 1$ rather then for $q_c\simeq q_{EA}$. In the following section we will discuss this limit which can also be used to obtain another series expansion.

\section{The $q_c=1$ case}
In this case the only possible couples of configurations are those in which a configuration is coupled with itself. It is easy to realize that the thermodynamics of such a constrained system is equal to that of a single system at temperature $T_{12}=1/(\beta_1+\beta_2)$. It is also natural that $q_1(x)=q_2(x)=p(x)$. By looking at equation (\ref{eqq_c}) we see that $q_c=1$ requires $\delta=\infty$. In this case the initial conditions on $m_1$ and $m_2$ read
\begin{equation} 
m_1(1,y_1,y_2)=m_2(1,y_1,y_2)=\tanh [\beta_1y_1+\beta_2y_2]
\end{equation}
By making the change of variables
\begin{equation}
S=\frac{\beta_1y_1+\beta_2y_2}{\beta_1+\beta_2}\ ,\ \ \ \ D=y_2-y_1\ ,
\end{equation}
the SP equations (\ref{SP1},\ref{SP2},\ref{SP3}) reduce to those of a single system at temperature $T_{12}$, as expected.
The case $T_1=T_2$ has been recently discussed in \cite{BFM}.
It is interesting to notice that the constrained model with $q_c=1$ has a critical temperature whose value is exactly twice as the critical temperature of the free model. Therefore if we consider a couple of constrained systems such that $T_1 \geq T_c$, $T_2 \geq T_c$ but $T_{12}\leq T_c$ we shall have a  phase transition from a paramagnetic replica-symmetric (RS) phase to a RSB phase by changing the value of the constraint $q_c$. 
For completeness we report the paramagnetic RS solution of the SP equations:
\begin{equation}
q_1(x)=q_2(x)=p(x)=0\ \ \ \  p_d={1 \over \beta_1\beta_2}\  {\rm arctanh}\  q_c
\end{equation}
Notice that $p_d$ is divergent for $q_c\rightarrow 1$; therefore, according to (\ref{PEPSI}), the derivative of the free-energy is divergent too. This result is valid also in the RSB phase.
For $T_1=T_2=T$, $1\leq T\leq 2$ the critical value of the constraint $q_c $ turns out to be  $q_{{\rm crit}}(T)=T-1$. For $q_c$ near $q_{{\rm crit}}$ the order parameter is small and we solved the equations perturbatively in $q_c-q_{	{\rm 	crit}}$. We also obtained  a perturbative expansion for $T_1\simeq 2$, $T_2\simeq 2$ and $q_c\simeq 1$. Indeed, in this region the three functions $q_1(x)$, $q_2(x)$ and $p(x)$ are small since the value $q_c=1$ corresponds to a free system at a temperature near $T_c=1$.

\begin{figure}[htb]
\begin{center}
\includegraphics{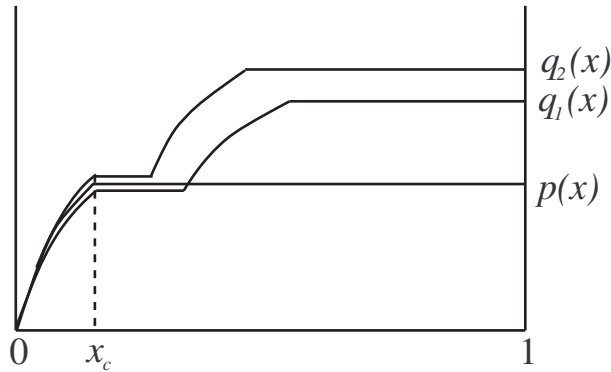}
\caption{The solutions discussed in \cite{rizzo1}. The real solutions are expected to pick up small corrections induced by chaos. The actual difference between the three functions in the small-$x$ region is much smaller than in the figure.}
\label{figure1}
\end{center}\end{figure}

\section{The $\beta_1\neq \beta_2$ case}

In \cite{rizzo1} an analytical picture for the absence of chaos in spin-glasses was proposed based on the possible existence of specific solutions of the SP equations (\ref{SP1},\ref{SP2},\ref{SP3}).
 The structure of these solutions is depicted in figure (\ref{figure1}): in the
small-$x$ region the three functions $q_{1}(x),q_{2}(x)$ and
$p(x)$ are all different till they reach the point $x_{c}$ where
$p(x)=p_{d}=q_{c}$; then for $x$ greater than $x_{c}$ $p(x)$
remains constant while $q_{1}(x)$ and $q_{2}(x)$ after an
intermediate plateau are connected continuously to the
corresponding free Parisi solutions. These solutions may exist for
values of the constraint from zero to a maximum value where the
two plateaus of the function at the higher temperature merge ($T_1>T_2$ in figure (\ref{figure1})). 
These solutions reduce to eq. (\ref{soliso}) in the limit $T_1\rightarrow T_2$ and are intrinsically non-chaotic due to the relation $\delta=0$ \cite{rizzo1}.
However, their existence must be checked explicitly by attempting to solve the SP equations (\ref{SP1},\ref{SP2},\ref{SP3}) with the ansatz $\delta=0$.  In \cite{rizzo1} this has been checked up to the fifth order in a perturbative expansion near the critical temperature, {\it i.e.} it has been found that $\Delta F$ is a quantity smaller than the fifth order.

Recently it has been proven that these solutions exist in the spherical spin-glass models with multi-$p$ spin interaction \cite{rizzo2}. In that context it has been shown that they imply ultrametricity between the equilibrium states at different temperatures so that the Parisi tree is essentially the same at all temperatures. Furthermore, arguments have been advanced in order to show that the reference free energies of the clusters of states do not change with temperature, much as in the generalized random-energy model. 
Perfect ultrametricity is a special property of those models due to the fact that the functions $q_1(x)$, $q_2(x)$ and $p(x)$ are all equal in the small-$x$ region. In general, however, we believe that the physical interpretation of the solutions proposed in \cite{rizzo1} is similar. Indeed, we think that the non-chaotic solutions proposed in \cite{rizzo1} are the mathematical implementation of the old idea of a unique Parisi tree which bifurcates when lowering the temperature plus the additional hypothesis that the ordering of the reference free energies of the clusters of states does not change when lowering the temperature. It would be interesting to check whether the logical equivalence between these solutions and the previous picture can be established explicitly through the cavity method as was done for the Parisi solution \cite{MPV}.

\subsection{Perturbative Expansion}
In order to deal with the SP equations (\ref{SP1},\ref{SP2},\ref{SP3}) of the SK model we resorted to a  high-order computer-assisted perturbative expansion using the methods we recently applied to the Parisi solution \cite{rizcris}.  
The procedure is conceptually straightforward: one tries to solve the SP equations with the non-chaotic ansatz of figure (\ref{figure1}) assuming that near the critical temperature the functions $q_1(x)$, $q_2(x)$, $p(x)$ are small and the $x$ regions where they vary are also small. Therefore one expands all the quantities of interest in series of $y$, $x$,  $\tau_1$, $\tau_2$ and $q_c$, ($\tau_i=1-T_i$); the first is of order $1/2$, the others are of order 1; the function $P(x,y)$, which is singular, can be treated considering its $y$ cumulants.
This is essentially the same computation that was  carried on in \cite{rizzo1} by hand up to the fifth order of the free-energy. It was found that up to that relatively high order the SP equations admit of a non-chaotic solution with $\delta=0$ for values of $q_c$ between $0$ and some maximum value near the self-overlap of the states at the higher temperature.  
Instead we have found that at ninth order in the free energy it is impossible to find a solution of the SP equation (\ref{SP1},\ref{SP2},\ref{SP3}) with $\delta=0$: {\em the non-chaotic solutions don't exist in the SK model}. Accordingly there must be a free-energy cost in imposing the constraint, which one expects to be of ninth order. 

Since off-equilibrium solutions must have $\delta\neq 0$, things become very complicated;  indeed, as explained in \cite{rizzo1}, when $\delta\neq 0$ the three functions $q_1(x)$, $q_2(x)$ and $p(x)$ are coupled in the whole small-$x$ region, {\it i.e.} if one of them is varying in some interval of $x$ the others must vary too. 
We guess that off-equilibrium solutions have the same structure of the non chaotic solutions of figure (\ref{figure1}) plus small corrections of order $\delta$, which is expected to be of the seventh order. Therefore we have tried to solve the SP equations (\ref{SP1},\ref{SP2},\ref{SP3}) with a complicated ansatz made up of five $x$ regions on each of whom the functions $q_1(x)$, $q_2(x)$ and $p(x)$ have a different expansion. Unfortunately, in spite of the very high number of parameters of this ansatz, we have not been able to solve the SP equations (\ref{SP1},\ref{SP2},\ref{SP3}) explicitly .
Therefore, in order to estimate the free-energy cost we plugged the same ansatz directly into equations (\ref{F}, \ref{eqF}) and extremized the free energy with respect to the variational parameters. We obtained the following estimate for the free-energy cost of constraining two copies of a SK spin-glass at temperature $T_1=1-\tau_1$ and $T_2=1-\tau_2$ to have a fixed overlap $q_c$:
\begin{equation}
\Delta F={12 \over 35}\ |q_c|^7(\tau_1-\tau_2)^2
\label{DF}
\end{equation}
As expected, this quantity turns out to be of the ninth order. Let us comment that although in principle this must be considered an upper bound for  the real value, it is most likely exact. 
Indeed, in \cite{rizcris} we observed that the order of the function $q(x)$ relevant for the exact value of the free energy is much lower than the one expected by power counting arguments; for instance if we maximize the free energy at order $n$, we are able to determine the order parameter at an order near $n/2$ rather than the order $n-2$ obtained by power counting.
Therefore even if we know that our ansatz is unable to reproduce the correct solution at the seventh  order, most likely this has no influence on the value of the free energy at the ninth order.

\begin{figure}[htb]
\begin{center}
\includegraphics{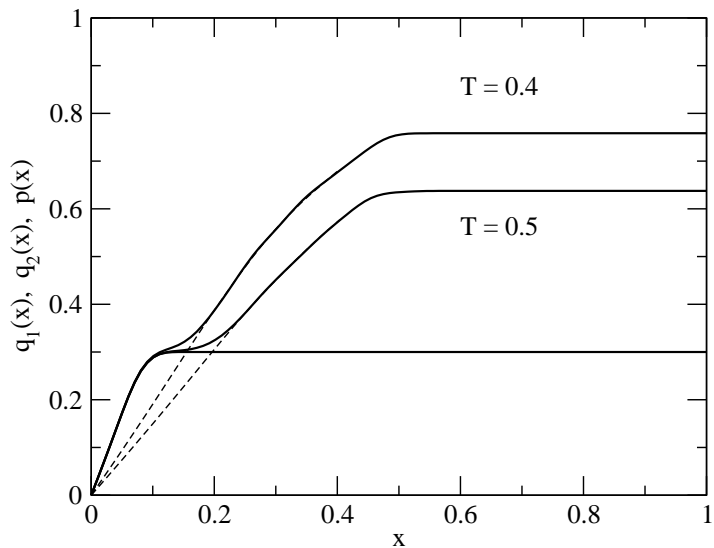}
\caption{The functions $q_1(x)$, $q_2(x)$ and $p(x)$ obtained numerically for $q_c=0.3$, $T_1=0.5$ and $T_2=0.4$. The dotted lines represent the corresponding free solutions. Within the numerical precision they coincide with those of figure (\ref{figure1}); the accuracy could be raised by the methods discussed in \cite{rizcris}}.
\label{figcris}
\end{center}\end{figure}

\subsection{Numerical Integration}
We also integrated the SP equations numerically {\it via} a pseudo-spectral code as we did in \cite{rizcris} for the standard Parisi solution.
Since the fields have three variables, time and memory required by the computation grow and  the obtainable precision decreases,  preventing us from seeing the chaos effect.
The advantage of this method, however, is that we can investigate the form of the solutions in a non-perturbative way. 
To obtain the three functions $q_1(x)$, $q_2(x)$ and $p(x)$, we applied the iterative scheme of \cite{NEM} to the SP equations (\ref{SP1},\ref{SP2},\ref{SP3}); we imposed at each iteration $p(x)<q_c$ and we set by hand  $\delta=0$; indeed $\delta$ is expected to be much smaller than the attainable precision. Consistently with this assumption, the free-energy cost computed integrating (\ref{F}) and (\ref{eqF}) turned out to be zero within the precision. 
Remarkably the three functions  $q_1(x)$, $q_2(x)$ and $p(x)$ have precisely the structure proposed in \cite{rizzo1} (cfr. figures (\ref{figure1}) and (\ref{figcris})). The chaos effect determines small corrections which should be seen at higher precision.
When $q_c$ is greater than some $q_{\max}$ the plateau of the function $p(x)$ stops at $q_{\max}$ irrespective of $q_c$, and in order to obtain higher values of $p(x)$ we should impose a finite value of $\delta$. In the isothermal case this happens when one goes out of equilibrium  for $q_c>q_{EA}$; when $T_1\neq T_2$ this effect marks a change in the behaviour of $\Delta F$ which is much larger for $q_c>q_{\max}$ (see discussion in the conclusions).
We believe that the chaos effect is beyond the attainable numerical precision of our code at the present, while improvements for the functions $q_1(x)$, $q_2(x)$ and $p(x)$ are at hand.

\section{Conclusions}
We settled the longstanding problem of chaos in temperature in the SK model: {\em chaos in temperature is present but it is an exceedingly small effect: $\Delta F$ is a ninth order quantity}.
The analytical picture for absence of chaos in temperature proposed in \cite{rizzo1} doesn't apply to the SK model, while it holds in the multi-$p$ spin spherical models \cite{rizzo2}; 
this is an important difference between these two models
whose behaviour has been often assumed to be similar. 

The effect is so weak that it can't be observed in present numerical simulations. Indeed, if we make an estimate for the function $P_N(q)$ through the equations (\ref{PNQ}) and (\ref{DF}) with parameters $T_1=.6$, $T_2=.4$ and $N=4096$, which are the temperatures and size investigated in the latest numerical work by Billoire and Marinari (BM) \cite{MARBILL2}, we find a completely flat function; therefore we  don't expect to see anything similar to the delta peak corresponding to the $N\rightarrow \infty$ limit. We notice that other corrections must be considered in order to explain those data; indeed, according to our results, a formula like $P_N(q)\propto  \exp[-N\Delta F]$ in any case would give a $P_N(q)$ with an absolute maximum in $q=0$.

Although we have not been able to determine exactly the off-equilibrium solutions corresponding to a non-zero constraint, we believe that they are similar to the non-chaotic solutions proposed in \cite{rizzo1} plus small correction of order $\delta=O(\tau^7)$; this is supported by the numerical integration of the SP equations (see fig. \ref{figcris}). As a consequence some of the features of the non-chaotic models discussed in \cite{rizzo2} are valid in the SK model at finite $N$.
For instance, we expect that the support of the function $P_N(q)$ has a maximum $q_{\max}$ located near the self-overlap $q_{EA}(T_{\rm great})$ of the system at the higher temperature. In the spherical models  this happens because $\Delta F=0$ for $q<q_{max}$ and $\Delta F>0$ for $q>q_{max}$; instead in the SK model $q_{\max}$ marks a crossover in the behaviour of the function $\Delta F$ which is ninth order for $q<q_{\max}$ while for $q>q_{\max}$ is much larger, most likely third order as in the isothermal case.
This is clearly seen in the numerical simulations of BM and also confirms their hypothesis that the approach to a delta function doesn't proceed by a splitting of the peak at $q_{\max}$ but rather by the emergence of a third peak in $q=0$.

\end{document}